\documentclass[11pt,a4paper]{article}
\usepackage{subfigure}
\usepackage{mathrsfs}
\usepackage{graphicx,color}
\usepackage{graphics,amsmath,amssymb,amsthm,amsfonts,amscd}
\def \t#1{\widetilde{#1}}

\numberwithin{equation}{section}
\numberwithin{lem}{section}

\title{Solutions to the complex Korteweg-de Vries equation:\\
Blow-up solutions and non-singular solutions}

\author{Ying-ying Sun$^1$,~~Juan-ming Yuan$^2$\footnote{E-mail: jmyuan@pu.edu.tw},~~Da-jun Zhang$^1$\footnote{Corresponding author. E-mail: djzhang@staff.shu.edu.cn}\\
\small\it{ $^1$Department of Mathematics, Shanghai University, Shanghai 200444,  P.R. China}\\
\small\it{$~^2$Department of Financial and Computational Mathematics, Providence University, Shalu, Taichung 433, Taiwan }
}

\begin{document}
\maketitle
\date{}
\begin{abstract}

In the paper two kinds of solutions are derived  for the complex Korteweg-de Vries
equation, including blow-up solutions and non-singular solutions.
We derive blow-up solutions from known 1-soliton solution and a double-pole solution.
There is a complex Miura transformation between the complex Korteweg-de Vries equation and
a modified Korteweg-de Vries equation.
Using the transformation, solitons, breathers and rational solutions to the complex Korteweg-de Vries equation
are obtained from those of the modified Korteweg-de Vries equation.
Dynamics of the obtained solutions are illustrated.

\vspace{0.5cm}


\end{abstract}
\section {Introduction}

The famous Korteweg-de Vries (KdV) equation,
\begin{equation}
 u_t +6uu_x + u_{xxx} = 0, \label{ckdv}
\end{equation}
is referred to as the complex KdV (cKdV) equation if $u=u(x,t)$ is a
complex-valued function. This equation provides an example admitting
wave blow-up in finite  time \cite{Birwir-SIAM-1987}, and also
severes as a model describing irrotational flows in a shallow water
channel \cite{LEvi-TMP-1994}. Blow-up of solutions of (complex) PDEs
has been largely investigated
numerically \cite{Hou-2008, Weid-2003,Yuan-2005,Wu-Yuan-2005}
, while exact solutions with blow-up are interesting as well.
In \cite{Bona-2009}, blow-up from 2-soliton solutions of the cKdV equation was
considered by   Bona and Weissler under the assumption of the space variable $x$ being  complex.
In \cite{Li-2009-blowup} Li made use of  Darboux Transformation to get exact
blow-up solutions to the cKdV equation \eqref{ckdv}.
In his treatment $x$ is kept real while the wave number $k$ is complex.
In fact,
\eqref{ckdv} can be viewed as a complex-valued integrable system,
and then it naturally admits $N$-complex soliton solutions. We will
show that solutions obtained in \cite{Li-2009-blowup} can also be
derived from 1-complex soliton.
Furthermore, the same technique will be applied to another solution of the cKdV equation
and derive possible blow-up points.

Besides, we will also derive exact solutions without singularity for
the cKdV equation, via Miura transformation, from solutions of the
modified KdV (mKdV) equation. In fact, there are two mKdV equations,
\begin{equation}
\mathrm{mKdV}^+:~~~ v_t + 6v^2{v_x} + v_{xxx} = 0, \label{mkdv}
\end{equation}
and
\begin{equation}
\mathrm{mKdV}^-:~~~ v_t - 6v^2{v_x} + v_{xxx} = 0. \label{mkdv-}
\end{equation}
These two equations exhibit different aspects \cite{PFE-mkdv-1974,Romanova-mkdv-1979,FGES-1991}.
On the real-value level, \eqref{ckdv} and \eqref{mkdv-} are related by the
well-known Miura transformation (MT) \cite{{Miura-JMP-1968}}
\begin{equation}
u=-v^2\pm v_x
\end{equation}
which provides an approach to the solutions of the KdV equation
\eqref{ckdv} from those of the mKdV$^-$ equation \eqref{mkdv-}. It
is hard to reverse the MT to get solutions of the mKdV$^-$ equation
(cf.\cite{FGES-1991}). In a recent paper \cite{Zhang-mkdv-2012}, we gave a complete
investigation on exact solutions of the mKdV$^+$ equation
\eqref{mkdv} in terms of Wronskians. We have
got solitons, breathers as well as non-singular rational solutions
(by means of a Galilean transformation) for \eqref{mkdv}. Note that
the MT between \eqref{ckdv} and \eqref{mkdv} is of complex-valued
form (cf.\cite{Buti-PS-1986})
\begin{equation}
u=v^2\pm iv_x. \label{mt}
\end{equation}
This means for the real-valued solution $v(t,x)$ of the mKdV$^+$ equation \eqref{mkdv},
the MT provides a complex-valued solution $u(t,x)$ to the cKdV
equation \eqref{ckdv}. This fact enables us to list non-singular
solutions to the cKdV equation \eqref{ckdv}, including solitons,
breathers and rational solutions.

The paper is arranged as follows. In Section 2, we  find out blow-up
solutions from a complex soliton solution and a double-pole solution. In Section 3, we derive
non-singular solutions to the cKdV equation. Dynamics of some
obtained solutions are also illustrated.


\section{Blow-up solutions}\label{Sec:2}

\subsection{Some exact solutions to the cKdV equation}

First, we note that no matter $u$ is real or complex in
\eqref{ckdv}, it is invariant under the Galilean transformation
\begin{equation}\label{GTs}
    u(x,t)\rightarrow \alpha+u(x+6\alpha t,t),~~~ \alpha\in \mathbb{C},
\end{equation}
which may provides non-zero asymptotics. Besides, if we set
\[u=u_1 + iu_2,  ~~u_1=\mathrm{Re} [u],~~u_2=\mathrm{Im} [u],\]
then the cKdV equation \eqref{ckdv} is split into
\begin{subequations}\label{u12}
    \begin{align}
    &  u_{1,t}+6u_1u_{1,x}-6u_2u_{2,x}+u_{1,xxx}=0, \label{u12a}\\
    &  u_{2,t}+6u_1u_{2,x}+6u_2u_{1,x}+u_{2,xxx}=0. \label{u12b}
    \end{align}
\end{subequations}

If we consider equation \eqref{ckdv} as a real-valued equation, then it
is a typical integrable system and it has variety of solutions. A
well-known result is that through the transformation
\begin{equation}
u=\frac{2(ff_{xx}-f_x^2)}{f^2} \label{trans-bil}
\end{equation}
equation \eqref{ckdv} is written into the bilinear form
\begin{equation}\label{bilinear}
    (D_t D_x+D_{x}^4)f \cdot f=0,
\end{equation}
where the $D$ is the Hirota's bilinear operator defined
as \cite{Hirota-2004}
\begin{equation}D^{m}_{t}D^{n}_{x}a(t,x)\cdot b(t,x)=(\partial _{t}-\partial
_{t'})^{m}(\partial _{x}-\partial _{x'})^{n}
a(t,x)b(t',x')|_{t'=t,x'=x}.
\end{equation}
$N$-soliton solution is expressed through \eqref{trans-bil} and
\begin{subequations}\label{so:N-ckdv}
\begin{equation}
f=\sum\limits_{\mu =0,1} {{\exp \biggl({\sum\limits_{j = 1}^{N}
{{\mu _j}{\xi _j} + \sum\limits_{1 \le j < l}^{N} {{\mu _j}{\mu
_l}{A _{jl}}} } }}}\bigg),
\end{equation}
where
\begin{equation}
   \xi_{j}=k_{j}x-k_{j}^{3}t+h_j,~
   e^{A_{jl}}=\Big(\frac{k_j-k_l}{k_j+k_l}\Big)^{2},~j,l=1,2,\cdots,N,
\end{equation}
\end{subequations}
$k_j,h_j$ are constants, and the summation of $\mu$ takes over all
possible combinations of $\mu _j~(j=1,2,\cdots,N).$ Particularly,
for 1-soliton ($N=1$) and 2-soliton ($N=2$) solutions, $f$
can be written respectively in the following
\begin{equation}\label{f-expand1}
f_{1}(x,t)=1+e^{\xi_{1}},
\end{equation}
and
\begin{equation}\label{f-2}
f_{2}(x,t)=1+e^{\xi_{1}}+e^{\xi_{2}}+\Big(\frac{k_1-k_2}{k_1+k_2}\Big)^{2}e^{\xi_{1}+\xi_{2}}.
\end{equation}

Then 1-soliton solution is given through the transformation \eqref{trans-bil}
as
\begin{equation}\label{sol-1-ckdv}
    u=2\frac{f_1f_{1,xx}-f^2_{1,x}}{f_1^2}=\frac{2k_1^{2}e^{\xi_1}}{(1+e^{\xi_1})^2}
\end{equation}
if $k_1,h_1\in \mathbb{R}$. If taking $k_j\in \mathbb{C}$ in \eqref{so:N-ckdv}, one will have
complex solutions to the cKdV equation \eqref{ckdv}.

\subsection{Blow-up solution from 1-solition}

The idea of finding blow-up points is as the following.
 We consider
$k_1$ and $h_1$ in $\xi_1$ are complex and for convenience we write
\[k_1=k_{11}+ik_{12},~h_1=h_{11}+ih_{12}, ~~k_{11},k_{12},h_{11},h_{12}\in \mathbb{R}.\]
In this case, $f_1$ is rewritten as
\begin{equation}
    f_1= 1+e^{\eta_1}(\cos{\theta_1}+i\sin{\theta_1})¨Û\label{f1-c}
\end{equation}
with
\begin{equation}
\eta_1=k_{11}x-k_{11}^3t+3k_{11}k_{12}^2t+h_{11},~~
\theta_1=k_{12}x+k_{12}^3t-3k_{12}k_{11}^2t+h_{12}.\label{xi1-c}
\end{equation}
Singularity of $u$ then appears when $f_1=0$, i.e.,
\begin{equation}
\left\{
\begin{array}{l}
 1+e^{\eta_1}\cos{\theta_1}=0,\\
 e^{\eta_1}\sin{\theta_1}=0,
\end{array}
\right.
\end{equation}
which holds if and only if
\begin{equation}
\left\{
\begin{array}{l}
 \eta_1=0,\\
 \theta_1=(2s+1)\pi,~~ s\in \mathbb{Z}.
\end{array}
\right.
\end{equation}
In more details, this is a linear system of $(x,t)$:
\begin{equation}\label{coefficient matrix}
  \left(
  \begin{array}{cc}
    k_{11} & k_{11}(3k_{12}^2-k_{11}^2) \\
   k_{12}& k_{12}(k_{12}^2-3k_{11}^2)\\
  \end{array}
\right) \left(\begin{array}{c}x\\t\end{array}\right)
=\left(\begin{array}{l}
-h_{11}\\
-h_{12}+(2s+1)\pi
\end{array}
\right).
\end{equation}
Blow-up points correspond to separated zeros of the above system,
which requires the coefficient matrix is non-singular, i.e.,
\[  \left|
  \begin{array}{cc}
    k_{11} & k_{11}(3k_{12}^2-k_{11}^2) \\
   k_{12}& k_{12}(k_{12}^2-3k_{11}^2)\\
  \end{array}
\right|\neq 0,\] say $k_{11}k_{12}(k_{11}^2+k_{12}^2)\neq 0$, or,
\begin{equation}
k_{11}k_{12}\neq 0. \label{cond-1blow}
\end{equation}
Separated blow-up points $\{(x,t)\}$ are then given by
\begin{subequations}\label{xt-1-blow}
\begin{align}
& x=-\frac{h_{11}}{k_{11}}+\frac{(3k_{12}^2-k_{11}^2)(k_{12}h_{11}
-k_{11}h_{12}+k_{11}\pi + 2k_{11}s
\pi)}{2k_{11}k_{12}(k_{11}^2+k_{12}^2)},\\
&
t=\frac{k_{11}h_{12}-k_{12}h_{11}-k_{11}(2s+1)\pi}{2k_{11}k_{12}(k_{11}^2+k_{12}^2)},
\end{align}
with $s\in\mathbb{Z}$.
\end{subequations}
The coordinates of the above blow-up points are the same as derived
in \cite{Li-2009-blowup}, up to the Galilean transformation
\eqref{GTs}.
Obviously, the blow-up points are on a straight line related to given $k$ and $h$, written as
\begin{equation}\label{xt-1ss}
    h_{11}+k_{11}x-k_{11}(k_{11}^2-3k_{12}^2)t=0.
\end{equation}

\begin{figure}[!h]\centering
\subfigure[]
{\label{fig:1-1} 
\includegraphics[width=2in,,height=1.6in]{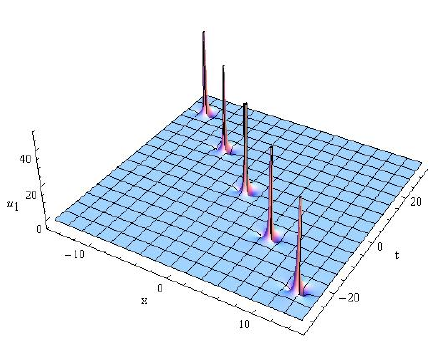}
}~~
\subfigure[]{
\label{fig:1-2} 
\includegraphics[width=2in,,height=1.7in]{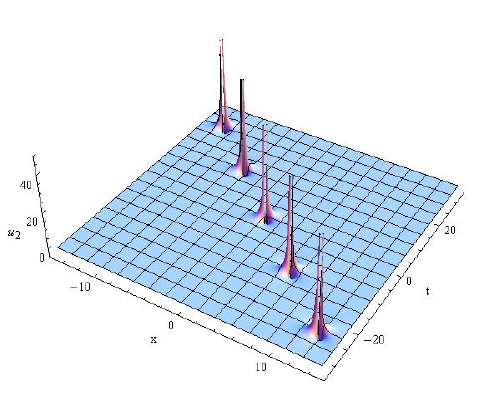}
}\\
\caption{Shape and motion of the blow-up solution given by
\eqref{1ss-blow-up}. (a) $u_1$ given by \eqref{1ss-blow-up-u1}.
(b) $u_2$ given by \eqref{1ss-blow-up-u2}.}
\label{fig:1} 
\end{figure}

As an example to illustrate,  we take
$k_{11}=\frac{1}{2},~k_{12}=-\frac{1}{2},~h_{11}=0,~h_{12}=\pi$,
then the blow-up points are $(x,t)=(-2s\pi, 4s\pi)$ which are on the line $2x+t=0$.
The corresponding blow-up
solution is
\begin{subequations}\label{1ss-blow-up-u}
\begin{equation}\label{1ss-blow-up}
u=\frac{ie^{\frac{1}{4}(1+i)(t+2x)}}{(e^{\frac{1}{4}(1+i)t+\frac{1}{2}x}-e^{\frac{1}{2}ix})^2}=u_1+iu_2,
\end{equation}
where
\begin{align}
&u_1=\frac{(-1 + e^{\frac{t}{2} + x})e^{\frac{1}{4} (t + 2 x)} \sin{\frac{1}{4}(t - 2 x)}}{[1 + e^{\frac{t}{2} + x} - 2 e^{\frac{1}{4} (t + 2 x)} \cos{\frac{1}{4}(t - 2 x)}]^2} ,\label{1ss-blow-up-u1}\\
&u_2=\frac{e^{\frac{1}{4} (t + 2 x)}(-2e^{\frac{1}{4} (t + 2 x)}+(1+e^{\frac{t}{2} + x})\cos{\frac{1}{4}(t - 2 x)})}{[1 + e^{\frac{t}{2} + x} - 2 e^{\frac{1}{4} (t + 2 x)} \cos{\frac{1}{4}(t - 2 x)}]^2},\label{1ss-blow-up-u2}.
\end{align}
The equation is depicted in Fig.\ref{fig:1}.
\end{subequations}

\subsection{Blow-up solution from double-pole  solution}\label{sec:3.2}

A double-pole solution to the KdV equation \eqref{ckdv} is given by
\eqref{trans-bil} with \cite{CZD-JPSJ-2002}
\begin{equation}
\label{f-double}
f(x,t)=1+(x-3k_1^2t)e^{k_1x-k_1^3t}-\frac{1}{4k_1^2}e^{2k_1x-2k_1^3t}.
\end{equation}
This corresponds to take the limit $k_2 \to k_1$ in \eqref{f-2}
with redefined
\[   e^{h_1}=\frac{1}{k_1-k_2},~~e^{h_2}=-\frac{1}{k_1-k_2}. \]

To analyze possible isolated blow-up points, we consider the case of
$k_1=ik_{12}$. Note that in this case $f$ can be written as
\begin{equation}
f=1-\frac{1}{4k_{12}^2}+(x+3k_{12}^2t+\frac{1}{2k_{12}^2}\cos
\theta)\cos \theta+ i (x+3k_{12}^2t
+\frac{1}{2k_{12}^2}\cos\theta)\sin {\theta}, \label{f-1/2}
\end{equation}
where
\[\theta=k_{12}x+k_{12}^3t.\]
Restricting $f=0$ and from the imaginary part one has either
\begin{subequations}
\begin{equation}\label{imf1.5-11}
\sin\theta=0,
\end{equation}
or
\begin{equation}\label{imf1.5-12}
2k_{12}^2(x+3k_{12}^2t) +\cos \theta=0.
\end{equation}
\end{subequations}
In the first case, \eqref{imf1.5-11} together with the real part of
$f$ requires
\begin{equation}\label{imf1.5-21}
\left\{
\begin{array}{l}
k_{12}x+k_{12}^3t=s\pi,\\
x+3k_{12}^2t=(-1)^{s+1} \big(1+\frac{1}{4k_{12}^2}\big),
\end{array}\right.
\end{equation}
with solution
\begin{equation}\label{f1.5-sol}
  x=\frac{(-1)^s + 4(-1)^s k_{12}^2 + 12k_{12}s\pi}{8k_{12}^2}, ~~~
  t=-\frac{(-1)^s + 4(-1)^s k_{12}^2 + 4k_{12}s\pi}{8k_{12}^4},
\end{equation}
which provides isolated blow-up points $(x,t)$. It is worth noting
that the points \eqref{f1.5-sol} of $k_{12}^2=1/4$ are not isolated,
which belongs to the second case, i.e., \eqref{imf1.5-12}.

The second case does not lead to any isolated blow-up points. In
fact, in the light of \eqref{imf1.5-12}, one has $k_{12}^2=1/4$ and
then \eqref{imf1.5-12} reduces to
\begin{equation}
G(t,x)=\frac{1}{2}\Bigl(x+\frac{3t}{4}\Bigr)+\cos
\Bigl(\frac{x}{2}+\frac{t}{8}\Bigr)=0. \label{G}
\end{equation}
Noting that
\[G_t(t,x)=\frac{3}{8}-\frac{1}{8}\sin\Bigl(\frac{x}{2}+\frac{t}{8}\Bigr)\neq 0,~~~\forall x,t\in \mathbb{R},\]
\eqref{G} determines an implicit function $t=t(x),~x\in \mathbb{R}$,
on which $f=0$. That means in the case of  \eqref{imf1.5-12} there
is no isolated blow-up point.

Let us sum up this subsection.
\begin{itemize}
\item{When $k_{11}=0$ and $k_{12}^2=\frac{1}{4}$, the corresponding solution $u$ is real,
and there always exists a moving singular point on the curve
\eqref{G}.}
\item{When $k_{11}=0$ and $k_{12}^2 \neq \frac{1}{4}$, there exist blow-up
points $(x,t)$ coordinated by \eqref{f1.5-sol} and located on two parallel straight lines
\begin{subequations}\label{xt-1.5}
\begin{align}
&12 k_{12}^4 t + 4 k_{12}^2x+ 4 k_{12}^2+1=0,\\
& 12 k_{12}^4 t + 4 k_{12}^2x- 4 k_{12}^2-1=0 .
\end{align}
\end{subequations}

 Solution $u$ of this
case is given by \eqref{trans-bil} with $f$ defined in
\eqref{f-1/2}.}
\end{itemize}

\begin{figure}[!h]\centering
\subfigure[]
{\label{fig:2-1} 
\includegraphics[width=2.2in,,height=1.7in]{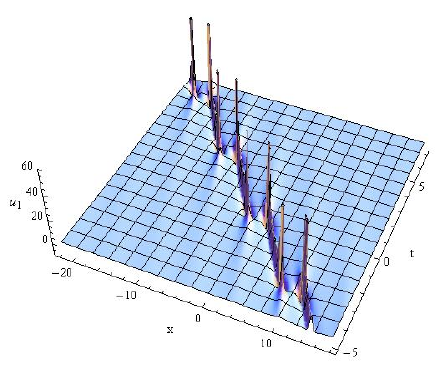}
}~~~
\subfigure[]{
\label{fig:2-2} 
\includegraphics[width=2.2in,,height=1.7in]{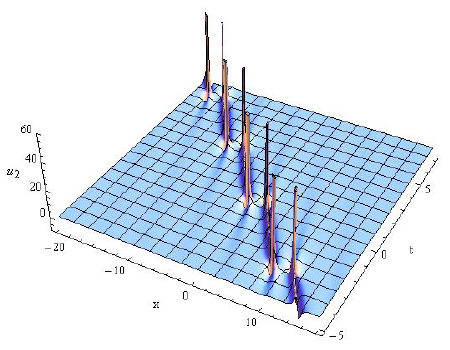}
}\\
\caption{Shape and motion of the blow-up solution given by
\eqref{k1=i}. (a) $u_1$ given by \eqref{u1--k1=i}.
(b) $u_2$ given by \eqref{u2--k1=i}.}
\label{fig:2} 
\end{figure}

For illustration we take $k_1=i$ and $u=u_{1}+i u_{2}$. In this case, we have
\begin{subequations}\label{k1=i}
\begin{align}
& u_1=\frac{B}{A^2},  \label{u1--k1=i} \\
& u_2=\frac{C}{A^2},\label{u2--k1=i}
\end{align}
\end{subequations}
where
\begin{equation*}
\begin{split}
A =&17 + 16(3t + x)^2 + 40(3t + x) \cos(t + x) + 8 \cos(2t + 2x),  \\
 B=&-8[(3 t + x) (20 \cos(3t + 3x) +
        (5 (85 + 16 (3 t + x)^2) \\
        &- 256 \sin(t + x))\cos(t + x)  +
        32 (3 t + x) (2 \cos(2t + 2x) - 5 \sin(t + x))]\\
    &-16[68 \cos(2t + 2x) + 25 \sin(t + x) +
        4 (8 + 33 (3t + x)^2 - 5 \sin(3t + 3x)))],\\
C=& 24[2(29 + 16(3t + x)^2) \cos(t + x) -
    8 \cos(
      3t + 3x) \\
        & + (3t +
       x) [80+   ((12 t + 4 x)^2-81  -
          8 \cos(2t + 2x)) \sin(t + x)] - 40 \sin(2t + 2x)] .
\end{split}
\end{equation*}
The corresponding blow-up points are
\begin{equation}
(x,t)=\Big(\frac{ (-1)^s 5 + 12 s \pi }{8},\, -\frac{(-1)^s 5 + 4s\pi}{8}\Big),~~s\in \mathbb{Z}.
\end{equation}
We plot $u_1$ and $u_2$ in Fig.\ref{fig:2}.

\newpage

\section {Non-singular solutions}\label{Sec:3}

By means of the MT \eqref{mt}, solutions of the cKdV equation \eqref{ckdv}
can be derived from those of the mKdV$^+$ equation \eqref{mkdv}.

\subsection {Exact solutions of the mKdV$^+$ equation}\label{Sec:3.1}

Let us list out solutions of the mKdV$^+$ equation \eqref{mkdv}
classified in \cite{Zhang-mkdv-2012}. We make use of Wronskians
$|\widehat{N-1}|$ and lower-triangular Toeplitz matrices set
$\widetilde{G}_N$ for which one can refer to Appendices \ref{sec:A}
and \ref{sec:B}.

\subsubsection{Solitons and breathers}

Soliton and breather solutions of the cKdV equation \eqref{ckdv} can be described by
\begin{subequations}
\label{nss}
\begin{equation}
v=2\left(\mathrm{arctan}\frac{F_2}{F_1}\right)_{x}
=\frac{-2(F_{1,x}F_2-F_1F_{2,x})}{F_2^2+F_1^2},
\label{nss-s}
\end{equation}
 where
\begin{equation}
f=f(\varphi)=|\widehat{N-1}|=F_1+iF_2,~~F_1=
\mathrm{Re}[f],~~F_2=\mathrm{Im}[f]. \label{nss-w}
\end{equation}
\end{subequations}
Wronskian vectors corresponding to different kinds of solutions are
the following.
\begin{itemize}
\item{
\textbf{For soliton solutions:}
\begin{subequations}
\label{nss-phi-soliton}
\begin{equation}
\varphi=\varphi^{[s]}_{N}=(\varphi_{1},  \varphi_{2}, \cdots,
\varphi_{N})^{T},
\end{equation}
with
\begin{equation}
\varphi^{}_{j}= a_{j}^+  e^{\xi_{j}}+ i a_{j}^-
 e^{-\xi_{j}}, ~\xi_{j}=k_{j}x-4k_{j}^{3}t+\xi_{j}^{(0)},~~
 a_{j}^+ , a_{j}^-, k_j, \xi_{j}^{(0)} \in \mathbb{R}.
\label{nss-phi-soliton-j}
\end{equation}
\end{subequations}
}
\item{\textbf{For limit solutions of solitons:}
\begin{subequations}
\label{nss-phi-soliton-lim}
\begin{equation}
{\varphi}=\varphi^{[ls]}_{N}(k_1)=\mathcal{A^+} \mathcal{Q}_{0}^{+}+
i \mathcal{A^-} \mathcal{Q}_{0}^{-},~~\mathcal{A}^{\pm} \in
\widetilde{G}_N(\mathbb{R}),
\end{equation}
with
\begin{equation}
 \mathcal{Q}^{\pm}_0=(\mathcal{Q}^{\pm}_{0, 0},
\mathcal{Q}^{\pm}_{0,1}, \cdots, \mathcal{Q}^{\pm}_{0, N-1})^T,~~
\mathcal{Q}^{\pm}_{0, s}=\frac{1}{s!}\partial^{s}_{k_1}e^{\pm
\xi_1},
\end{equation}
\end{subequations}
where $\xi_1$ is defined in \eqref{nss-phi-soliton-j}. }
\item{
\textbf{For breather solutions:}
\begin{subequations}
\label{nss-phi-breather}
\begin{equation}
{\varphi}=\varphi^{[b]}_{2N}=(\varphi_{11}, \varphi_{12},
\varphi_{21}, \varphi_{22}, \cdots, \varphi_{N1}, \varphi_{N2})^{T},
\end{equation}
with
\begin{eqnarray}
&& \varphi_{j1}= a_{j}  e^{\xi_j}+b_{j} e^{-{\xi}_j}, ~~
\varphi_{j2}=\bar{a}_{j}  e^{\bar{\xi}_j} -\bar{b}_{j}
e^{-{\bar{\xi}_j}}, \\
&& \xi_j=k_{j}x- 4 k_j^3 t+ \xi_{j}^{(0)},~~a_{j},
b_{j},\xi_{j}^{(0)}  \in \mathbb{C}. \label{nss-phi-breather-j}
\end{eqnarray}
\end{subequations}
}
\item{
\textbf{For limit solutions of breathers:}
\begin{subequations}
\label{nss-phi-breather-lim}
\begin{equation}
\varphi=\varphi^{[lb]}_{2N}(k_1)=(\varphi^+_{1, 1}, \varphi^-_{1,
2}, \varphi^+_{2, 1}, \varphi^-_{2, 2}, \cdots, \varphi^+_{N, 1},
\varphi^-_{N, 2})^{T},
\end{equation}
and the elements are given through
\begin{align}
\varphi^{+}&=(\varphi^+_{1, 1}, \varphi^+_{2, 1}, \cdots,
\varphi^+_{N,1})^{T}
=\mathcal {A}\mathcal{Q}^{+}_{0}+\mathcal{B}\mathcal {Q}^{-}_{0},\\
\varphi^{-}&=(\varphi^-_{1, 2}, \varphi^-_{2, 2}, \cdots,
\varphi^-_{N, 2})^{T} =\bar{\mathcal
{A}}\bar{\mathcal{Q}}^{+}_{0}-\bar{\mathcal{B}}\bar{\mathcal{Q}}^{-}_{0},
\end{align}
where $\mathcal{A}, \mathcal{B} \in \widetilde{G}_N(\mathbb{C})$,
\begin{equation}
 \mathcal{Q}^{\pm}_0=(\mathcal{Q}^{\pm}_{0, 0},
\mathcal{Q}^{\pm}_{0,1}, \cdots, \mathcal{Q}^{\pm}_{0, N-1})^T,~~
\mathcal{Q}^{\pm}_{0, s}=\frac{1}{s!}\partial^{s}_{k_1}e^{\pm
\xi_1},
\end{equation}
\end{subequations}
and $\xi_1$ is defined in \eqref{nss-phi-breather-j}. }
\item{
\textbf{Mixed solutions:}

Mixed solutions can be obtained by arbitrarily combining the above
vectors to be a new Wronskian vector. For example, take
\begin{equation}
\varphi=\left(\begin{array}{c}
\varphi^{[s]}_{N_1}\\
\varphi^{[ls]}_{N_2}(k_{N_1+1})
\end{array}
\right),
\end{equation}
then the related solution corresponds to the interaction between
$N_1$-soliton and a $(N_2-1)$-order limit-soliton solutions.}
\end{itemize}

\subsubsection{Rational solutions}

To get rational solutions, one has to make use of the Galilean
transformation
\begin{equation}
v(x,t)=v_{0}+V(X,t), ~~X=x-6{v_0}^{2}t,\label{GT}
\end{equation}
by which the mKdV$^+$ equation \eqref{mkdv} is transformed to a
mixed equation
\begin{equation}
{V_t}+ 12{v_0}V{V_X} + 6{V^2}{V_X} + {V_{XXX}} = 0,\label{kdv-mkdv}
\end{equation}
where $v_0$ is a real parameter. \eqref{kdv-mkdv} admits non-trivial
rational solutions which can be transformed back to those for the
the mKdV$^+$ equation \eqref{mkdv}. These rational solutions to
\eqref{mkdv} are given by
\begin{subequations}
\label{rss}
\begin{equation}
v(x,t)=v_{0}-\frac{2(F_{1,X}F_2-F_1 F_{2,X})}{F_2^2+F_1^2},~~
X=x-6v_{0}^{2}t,~~v_0\neq 0\in \mathbb{R}, \label{rss-s}
\end{equation}
where
\begin{equation}
f=f(\psi)=|\widehat{N-1}|=F_1+iF_2,~~F_1=
\mathrm{Re}[f],~~F_2=\mathrm{Im}[f], \label{rss-w}
\end{equation}
\end{subequations}
and the Wronskian is composed by
\begin{subequations}
\begin{equation}
\psi =(\psi_1,\psi_2,\cdots,\psi_N)^T,
\end{equation}
with
\begin{equation}
\psi_{j+1}= \frac{1}{(2j)!}\frac{\partial ^{2j} }{{\partial
k_1}^{2j}}\varphi_{1}\,\bigr|_{k_1=0},~~(j=0,1,\cdots,N-1),
\end{equation}
and
\begin{equation}
\varphi_{1}=\sqrt{2v_{0}+2ik_{1}}\,e^{\eta_1}+\sqrt{2v_{0}-2ik_{1}}\,
e^{-\eta_1}, ~~\eta_1=k_{1}X-4k_{1}^{3}t,~~X=x-6v_{0}^{2}t,~~ k_1\in
\mathbb{R}.
\end{equation}
\end{subequations}

\subsection{Miura transformation and solutions to the cKdV equation}

There is the following complex Miura transformation
\begin{equation}
u=v^2\pm iv_x \label{mt-c}
\end{equation}
to relate the cKdV equation \eqref{ckdv} and real mKdV$^+$ equation
\eqref{mkdv} through
\begin{equation}
   u_t + 6uu_x + u_{xxx} = (2v \pm i\partial_x)(v_t + 6v^2{v_x} + v_{xxx}).
\end{equation}
Noting that in \eqref{mt-c} $u=u_1+i u_2$ but $v$ is real, solutions
to the cKdV equation \eqref{u12} are then given as the following
\begin{equation}
u_1=v^2,~~~ u_2=\pm v_x, \label{u-v}
\end{equation}
where $v$ is a solution to the mKdV$^+$ equation \eqref{mkdv} and
has been listed out in Sec. \ref{Sec:3.1}. Using \eqref{u-v}, the equations \eqref{u12a}
and \eqref{u12b} can be transformed into
\begin{subequations}\label{v12}
    \begin{align}
    &  v_{t}+6v^2v_{x}+v_{xxx}=0, \label{v12aa}\\
    & (v_{t}+6v^2v_{x}+v_{xxx})_x=0,  \label{v12ab}
    \end{align}
\end{subequations}
then we can see the equation\eqref{v12aa} shares all its solutions
with the equation\eqref{v12ab}.

\eqref{u-v} also provides a possible transformation to bilinearize
the cKdV equation \eqref{u12}. Taking
\begin{equation}
u_1=-\biggl[\biggl( \ln\frac{\bar{f}}{f}\biggr)_x\biggr]^2,~~~
u_2=\pm i \biggl( \ln\frac{\bar{f}}{f}\biggr)_{xx}
\end{equation}
then we find both \eqref{u12a} and \eqref{u12b} can be bilinearized
as
\begin{subequations}
\begin{align}
(D_t+D_{x}^{3})\bar{f}\cdot f & =0,\label{eq4a}
\\
D_{x}^{2}\bar{f}\cdot f & =0,\label{eq4b}
\end{align}
\label{bilinear-kdv-mkdv}
\end{subequations}
which is nothing but the bilinear mKdV$^+$ equation.


\subsection{Examples and illustration}

Now let us have a look at some examples of solutions of the cKdV
equation \eqref{ckdv} or \eqref{u12}.
\begin{itemize}
\item{
\textbf{One-soliton solution:}
\begin{subequations}
\label{1ss-u1}
\begin{align}
& u_1= \frac{{16a{{_{\rm{1}}^{\rm{ + }}}^2}a{{_{\rm{1}}^{\rm{ -
}}}^2}{k_{\rm{1}}}^2}}{{{{(a{{_{\rm{1}}^{\rm{ - }}}^2}{e^{8k_1^3t -
2{k_1}x}} + a{{_1^{\rm{ + }}}^2}{e^{ - 8k_1^3t + 2{k_1}x}})}^2}}},
\\
& u_2=\pm\frac{{ 8a_1^{\rm{ + }}a_1^{\rm{ - }}k_1^2{e^{8k_1^3t{\rm{
+ }}2{k_1}x}}{\rm{(}}a{{_1^{\rm{ + }}}^{\rm{2}}}{e^{{\rm{4}}{k_1}x}}
- a{{_1^{\rm{ -
}}}^{\rm{2}}}{e^{{\rm{16}}k_1^3t}}{\rm{)}}}}{{{{{\rm{(}}a{{_1^{\rm{
+ }}}^{\rm{2}}}{e^{{\rm{4}}{k_1}x}}{\rm{ + }}a{{_1^{\rm{ -
}}}^{\rm{2}}}{e^{{\rm{16}}k_1^3t}}{\rm{)}}}^2}}},~~a_{1}^+ ,
a_{1}^-, k_1,  \in \mathbb{R}.
\end{align}
\end{subequations}
}

\item{
\textbf{One-breather solution:}
\begin{subequations} \label{1-bre-ss-u1}
\begin{align}
& u_1= 4{\Big[\Bigl(\mathrm{arctan}\frac{F_2}{F_1}\Bigr)_{x}}\Big]^2
=\frac{4(F_{1,x}F_2-F_1F_{2,x})^2}{(F_2^2+F_1^2)^2},
\\
& u_2=\pm2\Big(\mathrm{arctan}\frac{F_2}{F_1}\Big)_{xx}
=\pm2\Big(\frac{-F_{1,x}F_2+F_1F_{2,x}}{F_2^2+F_1^2}\Big)_x,
\end{align}

where

\begin{align}
F_1=&4k_{11}(a_{11}b_{11}+a_{12}b_{12})\cos(24k_{11}^2k_{12}t-8k_{12}^3t-2k_{12}x)\nonumber\\
~&+4k_{11}(a_{12}b_{11}-a_{11}b_{12})\sin(24k_{11}^2k_{12}t-8k_{12}^3t-2k_{12}x),
\label{F1-B}\\
F_2=&-2k_{12}e^{-2k_{11}(4k_{11}^2t+12k_{12}^2t+x)}\big[(b_{11}^2+b_{12}^2)e^{16k_{11}^3t}
+(a_{11}^2+a_{12}^2)e^{4k_{11}(12k_{12}^2t+x)}\big]. \label{F2-B}
\end{align}
\end{subequations}

}
\item{
\textbf{Rational solution:}
\begin{subequations} \label{1-rational-ss-u1}

\begin{align}
& u_1= \Bigl ( v_{0} - \frac{{4v_{0}}}{{1{\rm{ +
}}4{v_{0}^2}{\rm{(}}x - 6{v_{0}^2}t{\rm{)^2}}}}\Bigr)^2,
\\
& u_2=\pm\frac{{32{v_0^3}(x - 6{v_0^2}t)}}{{{{(1 + 4{v_0^2}{{(x -
6{v_0^2}t)}^2})}^2}}},~~~ v_0\neq 0 \in\mathbb{ R}.
\end{align}
\end{subequations}
}
\end{itemize}

These solutions are illustrated as the following, where the sign for $u_2$ we have taken ``$+$''.

\begin{figure}[!h]\centering
\subfigure[]
{\label{fig:3-1} 
\includegraphics[width=2in,,height=1.6in]{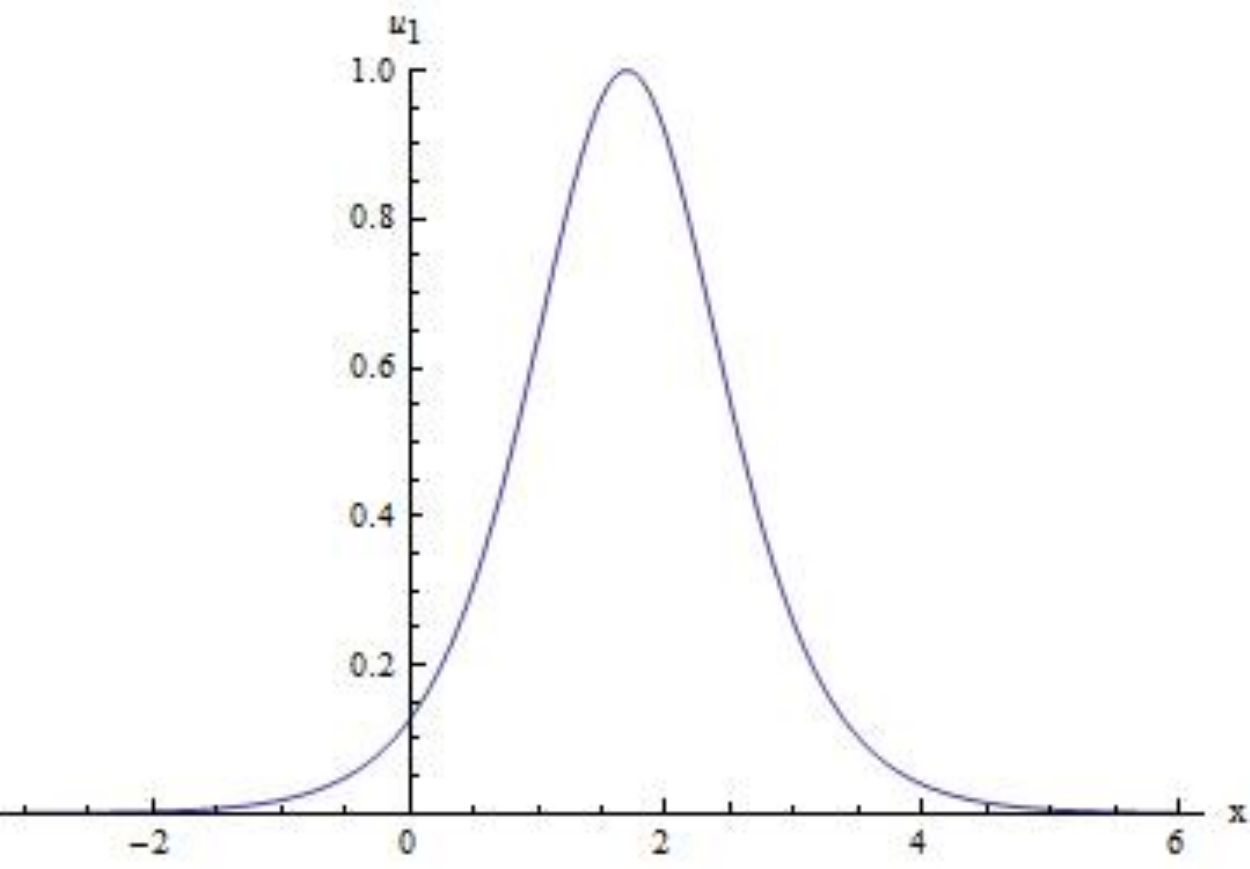}
}~~~
\subfigure[]{
\label{fig:3-2} 
\includegraphics[width=2.4in,,height=1.6in]{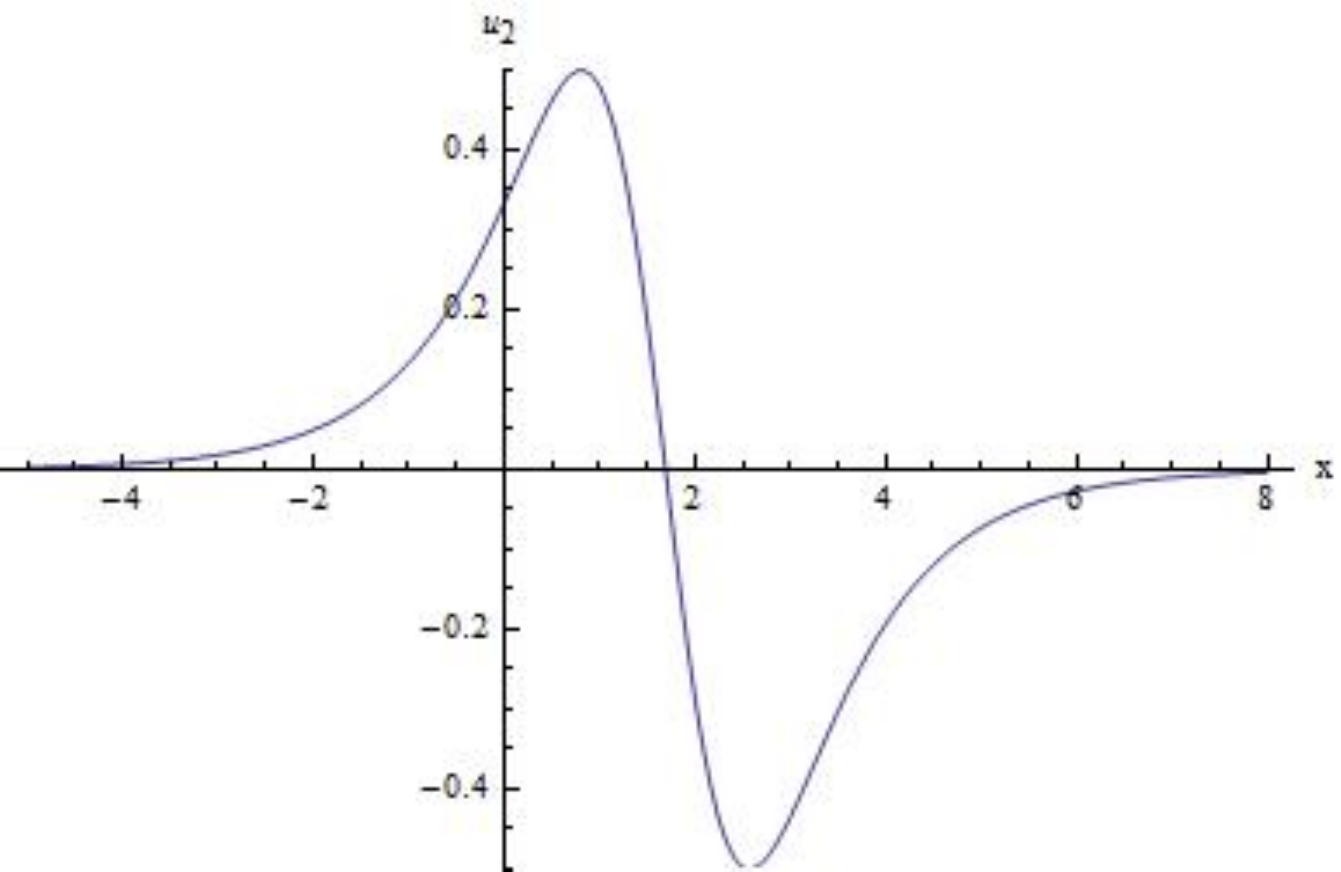}
}\\
\caption{Shape  of the one soliton solution given by
\eqref{1ss-u1} for $a_1^{\rm{+}}=1,~a_1^{\rm{-}}=2,~k_1=0.5$.}
\label{fig:3} 
\end{figure}

\begin{figure}[!h]\centering
\subfigure[]
{\label{fig:4-1} 
\includegraphics[width=2.1in,,height=1.6in]{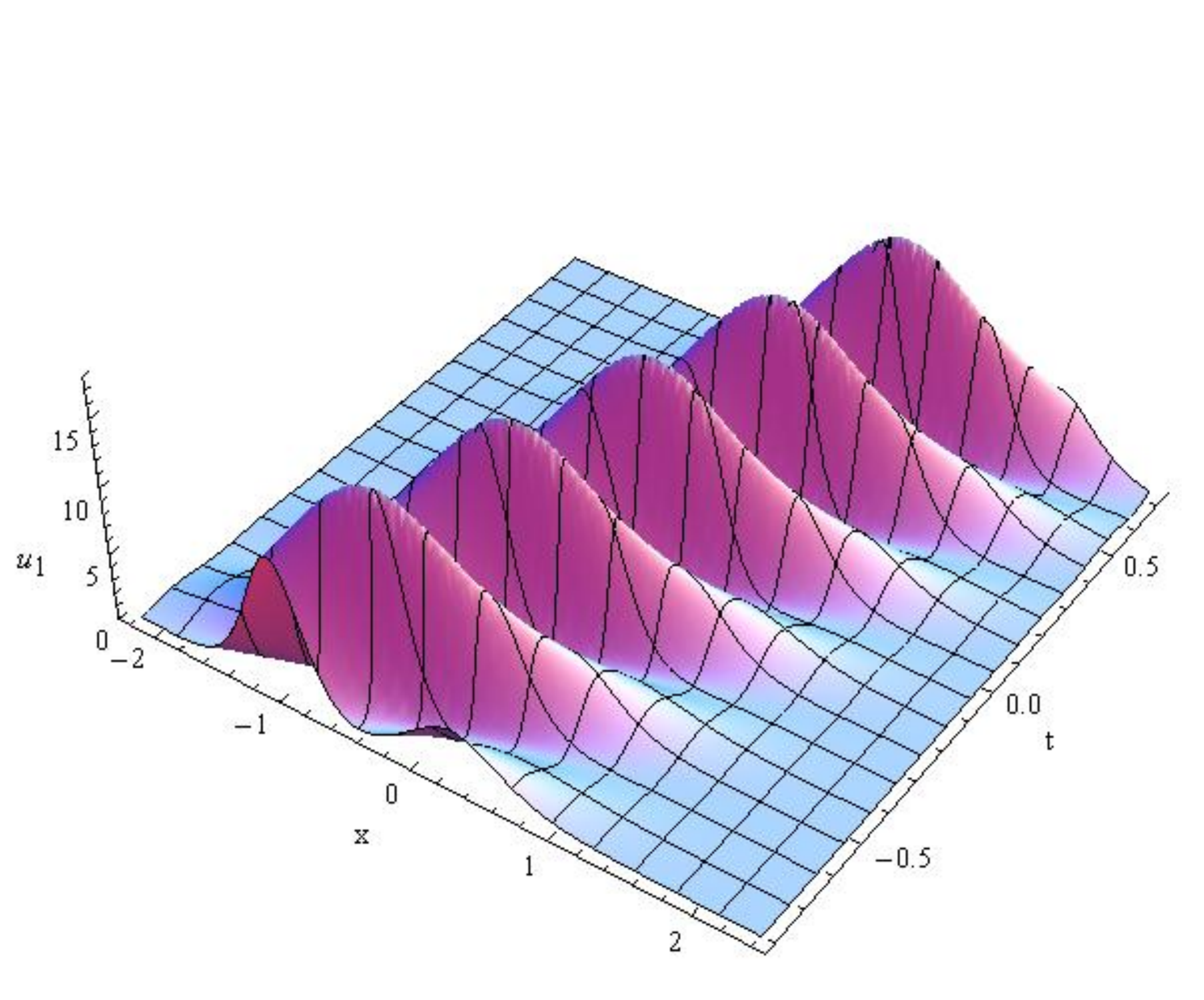}
}~~~
\subfigure[]{
\label{fig:4-2}
\includegraphics[width=2in,,height=1.6in]{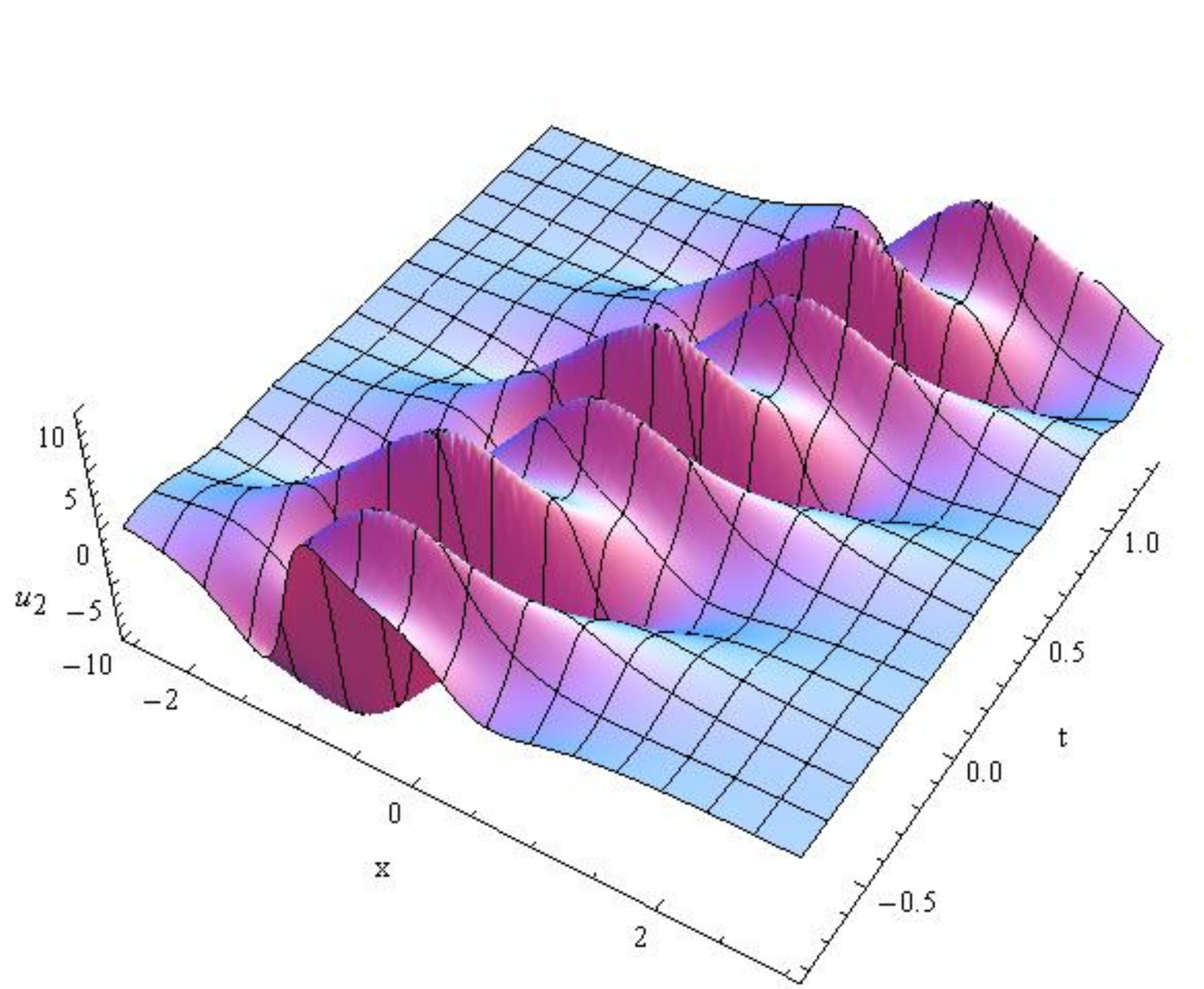}
}\\
\caption{Shape and motion of the one breather solution given by
\eqref{1-bre-ss-u1} for $k_1=1+ 0.5i,~a_1=1+i,~b_1= 2+ 0.5i$.}
\label{fig:4} 
\end{figure}

\begin{figure}[!h]\centering
\subfigure[]
{\label{fig:5-1} 
\includegraphics[width=2.7in,,height=1.6in]{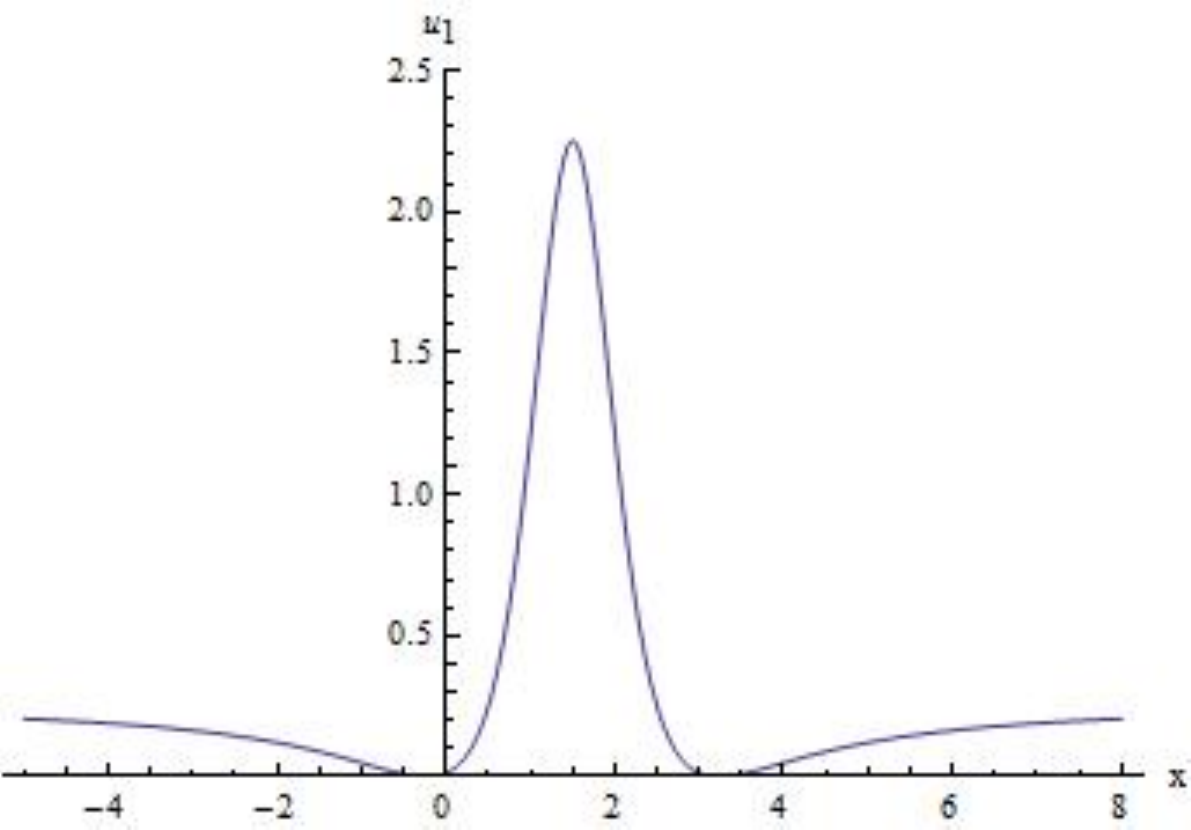}
}~~~
\subfigure[]{
\label{fig:5-2} 
\includegraphics[width=2.3in,,height=1.6in]{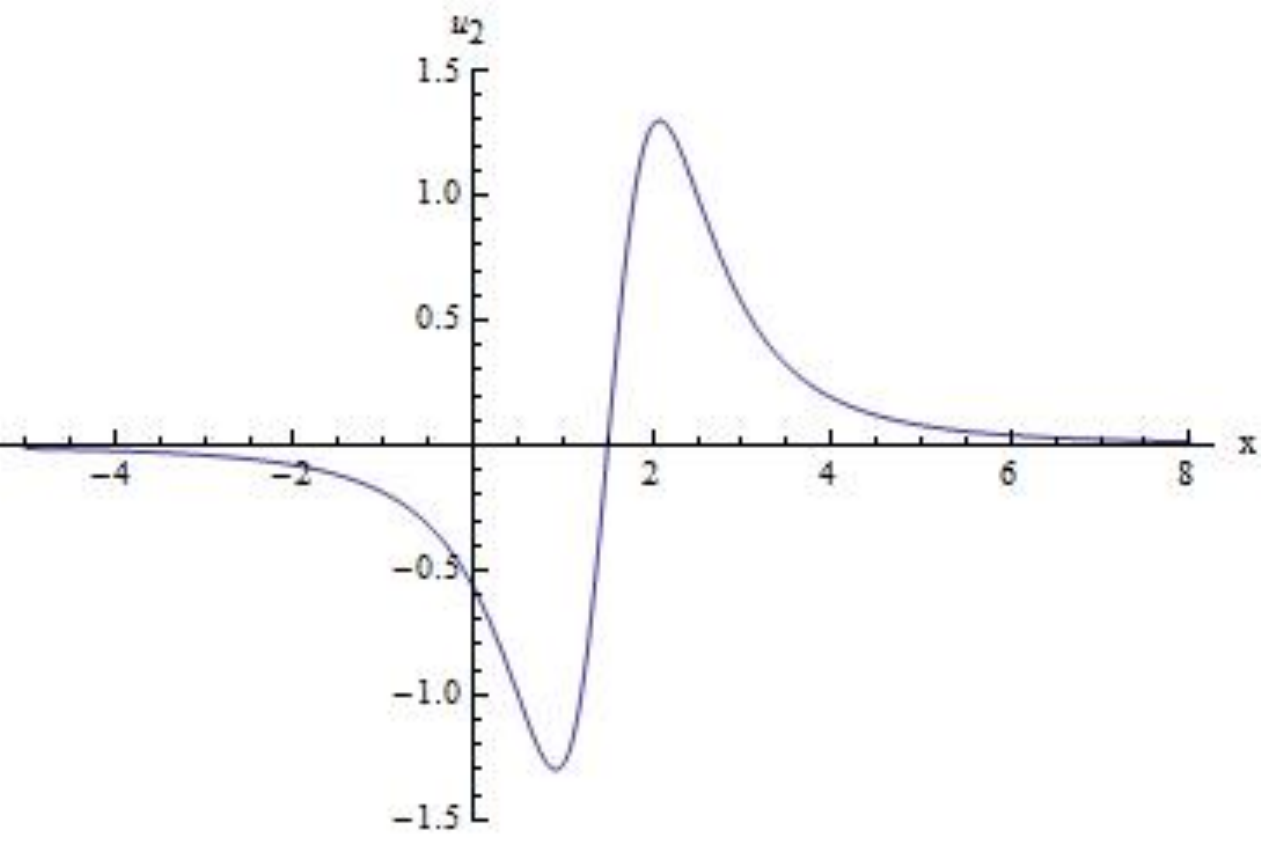}
}\\
\caption{Shape  of the rational solution given by
\eqref{1-rational-ss-u1} for $v_0=0.5$.}
\label{fig:5} 
\end{figure}

\section{Conclusions and discussions}

For the cKdV equation we have derived its two kinds of solutions: solutions with isolated blow-up points and
solutions without any singularities.
The real KdV equation admits variety of solutions.
By enlarging the field of wave numbers $\{k_j\}$ from $\mathbb{R}$ to $\mathbb{C}$,
we naturally get complex solutions for the cKdV equation.
We analyzed 1-soliton solution and the obtained isoslated blow-up points
are the same as those derived from Darboux transformation \cite{Li-2009-blowup}.
Besides, we considered a double-pole solution.
The solution admits isolated blow-up points located on two parallel straight lines,
which is a new feature for the waves with  blow-up.
For the solutions without any singularities,
we classified  them according to the classification
of solutions of the mKdV$^+$ equation.
The connection of the two equations is a complex Miura transformation.
Some numerical simulation results \cite{chenyong} cope with our exact solutions.
Besides, compared with the known exact solutions in \cite{zhangyi},
our bilinerlization for the cKdV equation and classification for its solutions are quite neat.

\begin{figure}[!h]\centering
\subfigure[]
{\label{fig:6-1} 
\includegraphics[width=2.2in,,height=1.8in]{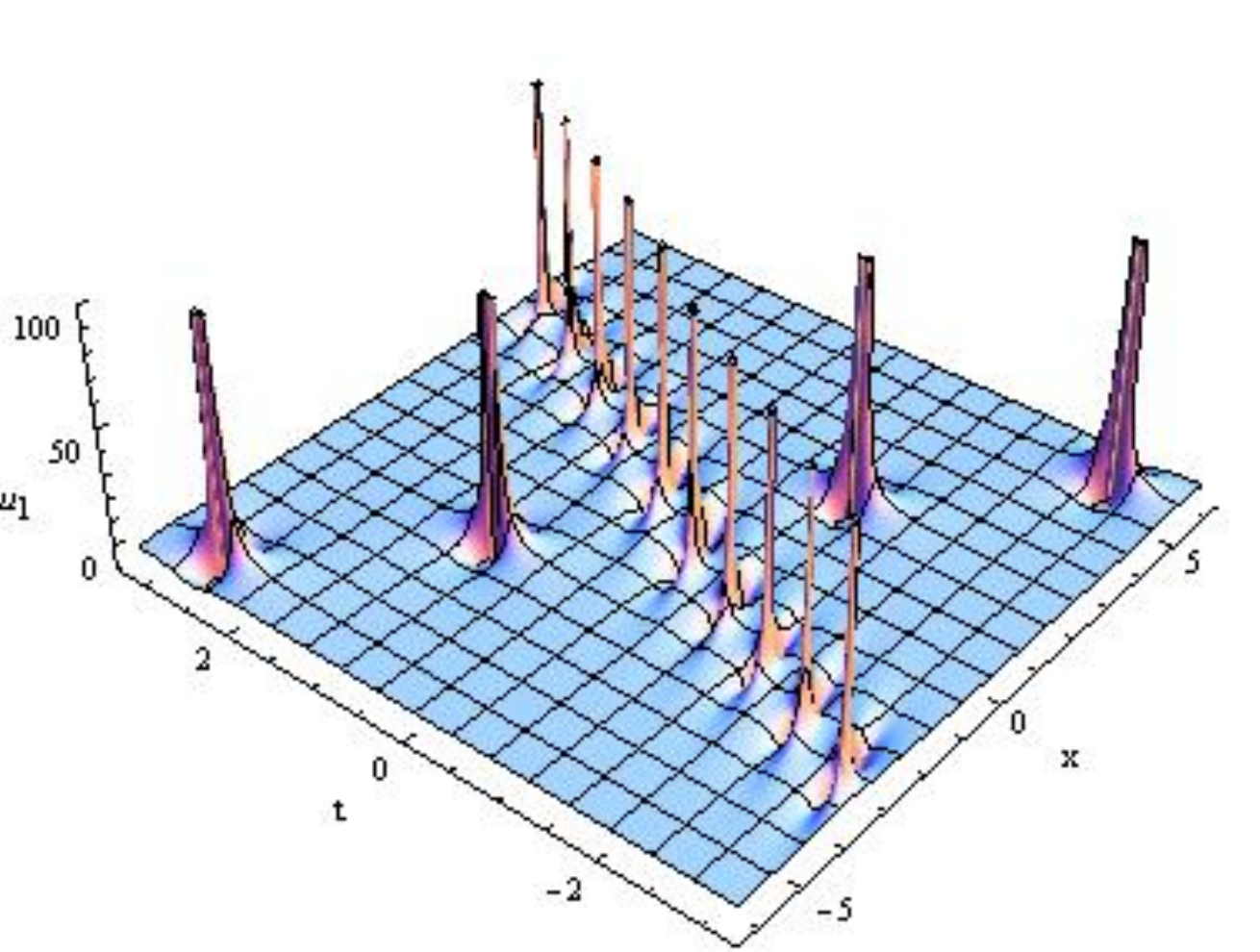}
}~~~
\subfigure[]{
\label{fig:6-2} 
\includegraphics[width=2.0in,,height=1.8in]{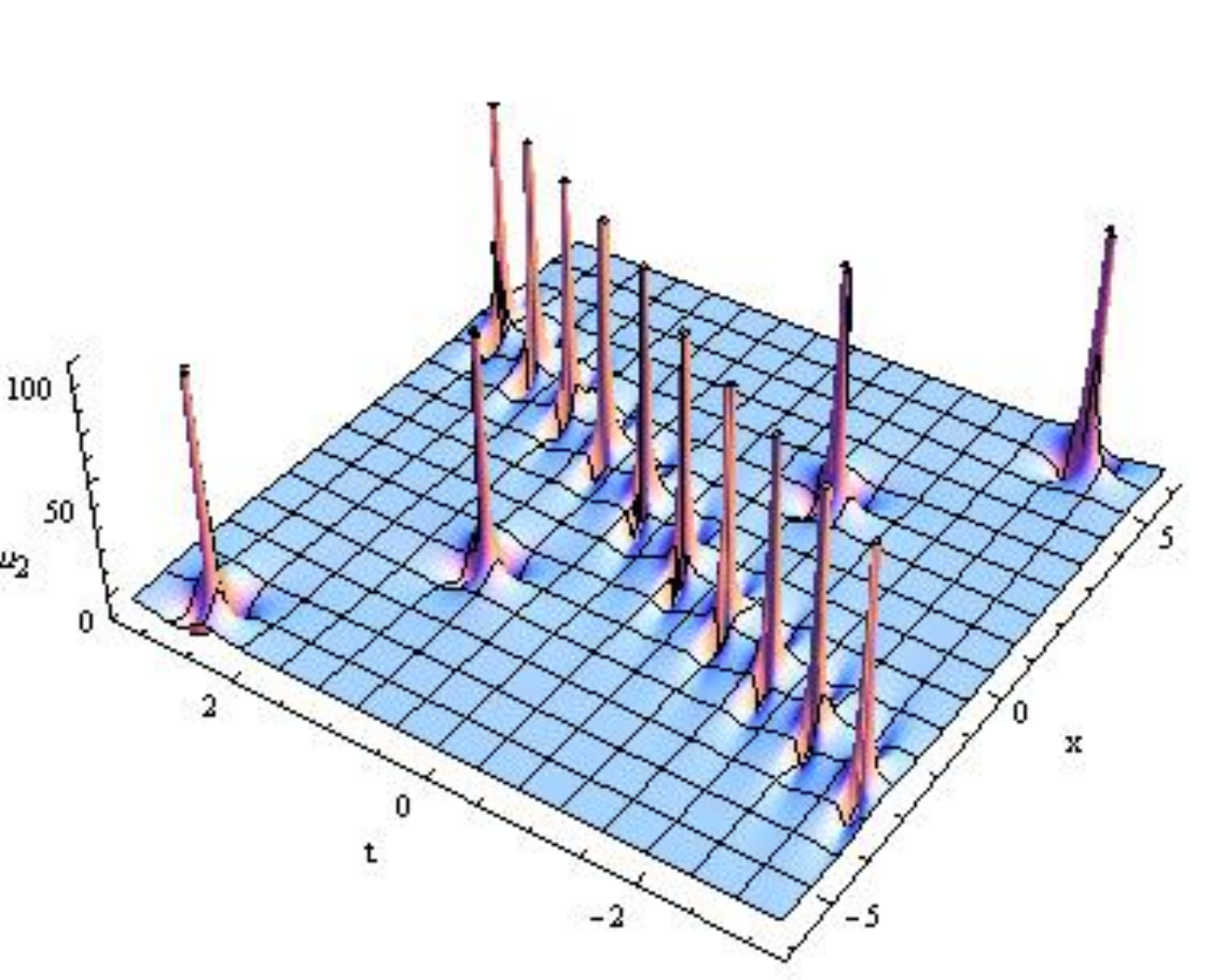}
}\\
\caption{Shape and motion of the 2-soliton solution given by \eqref{trans-bil}
with \eqref{f-2-2} for $k_1=2+i,~k_2 =1-i,~h_1=h_2=0$.
(a) $u_1=\mathrm{Re}[u]$. (b) $u_2=\mathrm{Im}[u]$.}
\label{fig:6} 
\end{figure}

For further discussion, let us back to the blow-up solutions.
For a solution with two complex wave numbers $k_1,k_2$, e.g. the 2-soliton solution
\eqref{trans-bil} with \eqref{f-2},
i.e.
\begin{equation}\label{f-2-2}
f_{2}(x,t)=1+e^{\xi_{1}}+e^{\xi_{2}}+\Big(\frac{k_1-k_2}{k_1+k_2}\Big)^{2}e^{\xi_{1}+\xi_{2}},
\end{equation}
the analysis will become more complicated.
For this moment we can not give a systematic analysis to write out
its isolated blow-up points, as we have done in our paper for the cases of 1-soliton
and double-pole solution.
However, for some special case we find the following dynamical picture (Fig.\ref{fig:6}),
which shows more straight lines for isolated blow-up points.

It is known that the cKdV equation is invariant under some transformations,
such as the Galilean transformation \eqref{GTs} and shift transformation $x\to x+x_0$.
Our isoslated blow-up points of 1-soliton case cope with the results of  \cite{Li-2009-blowup}
in light of the Galilean transformation \eqref{GTs}.
In  \cite{Bona-2009} Bona and Weissler
showed that 2-soliton solution of the cKdV equation admits blow-ups when $x_0\in \mathbb{C}$.
Their approach is quite different from ours.
Finally, let us revisit the double-pole solution generated from \eqref{f-double}.
For the real KdV equation such a solution has a singular point moving along two
logarithm curves (cf.\cite{zhang-tmp}),
while for the complex KdV equation, the blow-ups are located on two parallel straight lines, not logarithm curves.
Actually, \eqref{f-double} can be generalized to
\begin{equation}
\label{f-double-2}
f(x,t)=1+a(x-3k_1^2t+b)e^{k_1x-k_1^3t+h}-\frac{1}{4k_1^2}e^{2(k_1x-k_1^3t+h)},~~ k_1, a, b, h\in \mathbb{C},
\end{equation}
which can be derived either via Hirota's method(cf.\cite{CZD-JPSJ-2002}) or via some shift transformations $x\to x+x_0,~t\to t+t_0$.
Note that, as we have shown in Sec. 3.2, $f(x,t)$ given in \eqref{f-double} for the double-pole solution
can be viewed as a limiting result of \eqref{f-2-2} with $k_2\to k_1$.
It would be interesting to investigate the blow-up phenomenon of a general complex 2-soliton solution (eg. with complex $k_j$)
and its double-pole limit.

\vskip 20pt
\subsection*{Acknowledgments}
Yuan is partially supported by the National Science Council of the
Republic of China under the grant NSC 101-2115-M-126-002.
Yuan thanks Professors I-Liang Chern, Jyh-Hao Lee,
Yue Liu and Jiahong Wu for their encouragements.
Sun is supported by the Postgraduate Innovation Foundation of Shanghai University (No. SHUCX120112).
Zhang is supported by the NSF of China (No. 11071157), Project of ``The First-class Discipline of Universities in Shanghai''
and Shanghai Leading Academic Discipline Project (No. J50101).



\begin{appendix}

\section{Wronskians}
\label{sec:A}

A $N\times N$ Wronskian is defined as
\begin{equation}
W(\phi _{1},\phi _{2},\cdots ,\phi _{N})
=|\phi,\phi^{(1)},\cdots,\phi^{(N-1)}|=\left|
\begin{array}{cccc}
\phi _{1}^{(0)} & \phi _{1}^{(1)} & \cdots  & \phi _{1}^{(N-1)} \\
\phi _{2}^{(0)} & \phi _{2}^{(1)} & \cdots  & \phi _{2}^{(N-1)} \\
\vdots  & \vdots  &  \vdots & \vdots \\
\phi _{N}^{(0)} & \phi _{N}^{(1)} & \cdots  & \phi _{N}^{(N-1)}%
\end{array}
\right|,\label{Wro-def}
\end{equation}
where $\phi_{j}^{(l)}=\partial^l \phi_j/{\partial x}^l$ and
$\phi=(\phi_1, \phi_2,\cdots,\phi_N)^T$ is called the entry vector
of the Wronskian. Usually we use the compact
form\cite{Freeman-Nimmo-KP}
\begin{equation}
W(\phi)=|\phi,
\phi^{(1)},\cdots,\phi^{(N-1)}|=|0,1,\cdots,N-1|=|\widehat{N-1}|,
\label{wronskian}
\end{equation}
where $\widehat{N-j}$ indicates the set of consecutive
columns $0,1,\cdots,N-j$.
A Wronskian  provides simple forms for its derivatives
and this advantage admits direct verification of solutions that are expressed in terms of Wronskians.

\section{Lower-triangular Toeplitz matrices}
\label{sec:B}

A $N$th-order lower triangular Toeplitz matrix is a matrix  in the
following form
\begin{equation}
\mathcal{A}=\left(\begin{array}{cccccc}
a_0 & 0    & 0   & \cdots & 0   & 0 \\
a_1 & a_0  & 0   & \cdots & 0   & 0 \\
a_2 & a_1  & a_0 & \cdots & 0   & 0 \\
\cdots &\cdots &\cdots &\cdots &\cdots &\cdots \\
a_{N-1} & a_{N-2} & a_{N-3}  & \cdots &  a_1   & a_0
\end{array}\right)_{N\times N},~~~ a_j\in \mathbb{C}.
\label{A}
\end{equation}
All such matrices compose a commutative semigroup $\widetilde{G}_N(\mathbb{C})$ with identity
with respect to matrix multiplication and inverse,
and the set $G_N(\mathbb{C})=\bigl \{\mathcal{A} \bigl |~\bigr. \mathcal{A}\in \widetilde{G}_N(\mathbb{C}),~|\mathcal{A}|\neq 0 \bigr\}$
makes an Abelian group.

For more details, please refer to Ref.\cite{ZDJ-arxiv}.

\end{appendix}



\begin{thebibliography}{80}
\bibitem{Birwir-SIAM-1987} B. Birnir,
        An example of blow-up, for the complex KdV equation and existence beyond the blow-up, SIAM J. Appl. Math., 47(1987) 710-725.

\bibitem{LEvi-TMP-1994}D. Levi, Levi-Civita theory for irrotational water
        waves in a one-dimensional channel and the complex Korteweg-de Vries
        equation, Theor. Math. Phys., 99(1994) 705-709.


\bibitem{Hou-2008} T.Y. Hou and R. Li, Blowup or no blowup?
         The interplay between theory and numerics, Physica D, 237(2008) 1937-1944.


\bibitem{Weid-2003} J.A.C. Weideman, Computing the dynamics of complex singularities of
         nonlinear PDEs, SIAM J. Appl. Dyn. Syst., 2(2003) 171-186.


\bibitem{Yuan-2005} J.-M Yuan and J. Wu, The complex KdV equation with or without dissipation,
         Discr. Cont. Dyn. Syst. B, 5(2005) 489-512.
\bibitem{Wu-Yuan-2005} J. Wu and J.-M. Yuan, The effect of dissipation on solutions of the
         complex KdV equation,  Math. Comput. Simul., 69(2005) 589-599.

\bibitem{Bona-2009} J.L. Bona and F.B. Weissler,
         Pole dynamics of interacting solitons and blowup of complex-valued solutions of KdV,
         Nonlinearity, 22(2009) 311-349.

\bibitem{Li-2009-blowup} Y.C. Li,
         Simple explicit formulae for finite time blow up solutions to the complex KdV equation,
         Chaos, Solitons \& Fractals, 39(2009) 369-372.

\bibitem{PFE-mkdv-1974}T.L. Perel'man, A.Kh. Fridman and M.M. El'yashevich,
        A modified Korteweg-de Vries equation electrohydrodynamics,
        Zh. Eksp. Teor. Fiz., 66(1974) 1316-1323.

\bibitem{Romanova-mkdv-1979}N.N. Romanova,
        $N$-Soliton solution ``on a pedestal'' of the modified Korteweg-de Vries equation,
        Theor. Math. Phys., 39(1979) 415-421.

\bibitem{FGES-1991}F. Gesztesy, W. Schweiger and B. Simon,
        Commutation methods applied to the mKdV-equation,
        Trans. Amer. Math. Soc., 324(1991) 465-525.

\bibitem{Miura-JMP-1968}R.M. Miura,
        Korteweg-deVries equation and generalizations. I: A remarkable explicit nonlinear transformation,
        J. Math. Phys., 9(1968) 1202-1204.
\bibitem{Zhang-mkdv-2012} D.J. Zhang, S.L. Zhao, Y.Y. Sun and J. Zhou,
        Solutions to the modified Korteweg-de Vries equation, arxiv:1203.5851.

\bibitem{Buti-PS-1986} B. Buti, N.N. Rao and B. Khadkikar,
        Complex and singular solutions of KdV and MKdV equations,
        Phys. Scr., 34(1986) 729-731.
\bibitem{Hirota-2004}R. Hirota,
        \textit{The Direct Method in Soliton Theory}(in English),
        Cambridge University Press, 2004.

\bibitem{CZD-JPSJ-2002} D.Y. Chen, D.J. Zhang and S.F. Deng, Remarks on some solutions of soliton
         equations, J. Phys. Soc. Jpn., 71(2002) 2072-2073
\bibitem{chenyong} H.L. An and Y. Chen,
         Numerical complexiton solutions for the complex KdV equation by the homotopy perturbation method,
         Appl. Math. Comput., 203(2008) 125-133.
\bibitem{zhangyi} Y. Zhang, Y.N. L\"u, L.Y. Ye and H.Q. Zhao,
        The exact solutions to the complex KdV equation,
        Phys. Lett. A, 367(2007) 465-472.
\bibitem{zhang-tmp}D.J. Zhang, J.B. Zhang and Q. Shen,
         A limit symmetry of the KdV equation and its applications,
         Theore. Math. Phys., 163(2), (2010) 634-43.


\bibitem{Freeman-Nimmo-KP} N.C. Freeman and J.J.C. Nimmo,
         Soliton solutions of the KdV and KP equations: the Wronskian technique,
         {Phys. Lett. A}, {95}(1983) 1-3.


\bibitem{ZDJ-arxiv} D.J. Zhang, Notes on solutions in Wronskian form to soliton equations: KdV-type,
         arXiv:nlin.SI/0603008.







\end{thebibliography}
\end{document}